\UseRawInputEncoding
\documentclass[conference]{IEEEtran}
\IEEEoverridecommandlockouts
\usepackage{cite}
\usepackage{amsmath,amssymb,amsfonts}

\usepackage{graphicx}
\usepackage{textcomp}
\usepackage[a4paper,top=54pt,  left=37pt , right=37pt, bottom=113 pt]{geometry}
\usepackage{xcolor}

\begin{document}

\title{Single Image Super-resolution with a Switch Guided Hybrid Network for Satellite Images}

\author{1. Shreya Roy 2. Dr. Anirban Chakraborty \\ 
\emph{MTech(Research) student at Indian Institute of Science, Bangalore(1).}\\
\emph{Assistant Professsor at Indian Institute of Science, Bangalore(2)}\\  
\emph{shreyaroy@iisc.ac.in(1), anirban@iisc.ac.in(2)}
}

\maketitle
\thispagestyle{empty}

\begin{abstract}
The major drawbacks with Satellite Images are low resolution,
Low resolution makes it difficult to identify the objects
present in Satellite images.
We have experimented with several deep models available
for Single Image Superresolution on the SpaceNet dataset
and have evaluated the performance of each of them on the
satellite image data.
We will dive into the recent evolution of the deep models in
the context of SISR over the past few years and will present
a comparative study between these models.
\par  The entire Satellite image of an area is divided into equal-sized patches. Each patch will be used independently for training. These patches will differ in nature. Say, for example, the patches over urban areas have non-homogeneous backgrounds because of different types of objects like vehicles, buildings, roads, etc. On the other hand, patches over jungles will be more homogeneous in nature. Hence, different deep models will fit on different kinds of patches. In this study, we will try to explore this further with the help of a Switching Convolution Network. The idea is to train a switch classifier that will automatically classify a patch into one category of models best suited for it.

\end{abstract}

\section{Introduction}
Existing image Super-resolution algorithms can be classified into two categories:
\par
\label{sec:intro}
\textbf{Reconstruction Based:} Reconstruction-based algorithms require multiple
spatial/spectral/temporal low-resolution images of the same scene.
Learning-based approaches rely on prior information which can be
extracted from the existing dataset consisting of high and corresponding low resolution images.\par
\textbf{Learning-based Super-resolution:} Learning-based Super-resolution algorithms can be divided into three categories
regression, representation, and deep-learning-based algorithms.
\par
LR to HR mapping is a one to many mapping. So with deep learning approach we actually try to find the best mapping through which corresponds to the original HR image.
\par
In our work,we have experimented with several deep models available for Single Image Superresolution on the SpaceNet dataset and have evaluated the
performance of each of them on the satellite image data. We have found out two deep networks
DRLN and DBPN that have outperformed the present state of the art on the SpaceNet dataset in the
context of Single Image Super-resolution with Multi-scale Deep Residual Network for Satellite Image Super-Resolution Reconstruction \cite{xu2019multi}. Their result is present in Fig 0. In this experiment we have incorporated the idea of switch which will determine the deep network the image should be forwarded to. 
\par
We have tried to find out which deep model works better for the different types of regions like urban areas or rural areas, or areas with building or only field. For this we have referred the paper \cite{sam2017switching}'Switching Convolutional Neural Network for Crowd Counting' by Deepak Babu, Sam Shiv, Surya R. Venkatesh Babu. The idea is to train a switch which will divide the images into one of the categories of the deep models which will be more suitable to perform the super resolution  on the image patch. 
\par 
Also, to maintain the trade-off between computational time and the necessity of ensuring minute objects in the super-resolved image, we have decided to use the switch classifier.
\par
Single Image Super-resolution deep algorithms into 4 broad
Categories:
\subsection{A. Pre-upsampling Super-resolution:}
SRCNN , VDSR , DRRN, IRCNN, DNCNN are the examples under this category of
algorithm. In SRCNN Bicubic interpolation is done first to upsample to the desired
resolution and then 9×9, 1×1, 5×5 convolutions are performed to improve the image quality.
The 1×1 conv is used for non-linear mapping of the low-resolution
(LR) image vector and the high-resolution (HR) image vector.
\subsection{B. Post-upsampling Super-resolution:}
Examples of this kind are FSRCNN,ESPCN.  Five steps followed in FSRCNN involving more convolution layers are: 
\textbf{1. Feature Extraction:} Bicubic interpolation in previous SRCNN is replaced by 5x5 conv.\par
\textbf{2. Shrinking:} 1x1 conv is done to reduce the number of feature maps from d to s where
s is very very less than d. \par
\textbf{3. Non-Linear Mapping:} Multiple 3X3 layers are to replace a single wide one \par
\textbf{4. Expanding:} 1x1 conv is done to increase the number of feature maps from s to d. \par
\textbf{5. Deconvolution:} 9x9 filters are used to reconstruct the HR image. \par
\par
\subsection{C. Progressive Upsampling Super-
Resolution:}A few shortcomings of the previous architectures are addressed here. 
Bicubic interpolation is used to upscale an input LR image before going to network in pre upscaling SR architecture. 
However, this pre-upsampling step increases unnecessary computational cost and does not
provide additional high-frequency information for reconstructing HR images. 
Another drawback is optimize the networks with an L2 loss (i.e., mean squared error
loss). Since the same LR patch may have multiple corresponding HR patches and the L2
loss fails to capture the underlying multi-modal distributions of HR patches resulting in
the output images that are over-smoothed and inconsistent to human visual
perception on natural images. \par
\par Also, one step upsampling does not super-resolve the fine
structures well. When it comes to the learning mapping functions for large scaling factors (e.g., 8×), these single steps upsampling fail to capture the detail in a fine scale.
To address these issues, the deep Laplacian Pyramid Super-Resolution Network
(LapSRN)\cite{lai2017deep} was proposed to progressively reconstruct HR images in a coarse-to-fine
fashion. It consists of two branches Feature Extraction and Feature Reconstruction.
\par
Also, we must not forget to mention the generative model, the LAPGAN (introduced in 2016 by Denton \cite{denton2015deep}) is a architecture which is designed to synthesize diverse natural images
from random noise and sample inputs. On the other hand, the LapSRN predicts a particular HR image based on the given LR image and upsampling scale factor. The LAPGAN uses a cross-entropy loss function to encourage the output images to respect
the data distribution of the training datasets. In contrast, in LapSRN they use the Charbonnier
penalty function to penalize the deviation of the SR prediction from the ground truth HR images.
The LAPGAN upsamples input images before applying convolution at each level, while
LapSRN extracts features directly from the LR space and upscales images at the end of each
level which effectively alleviates the computational cost and increases the size of receptive fields. In
addition, the convolutional layers at each level in the LapSRN are connected through multi-channel
transposed convolutional layers. The residual images at a higher level are therefore predicted by a
deeper network with shared feature representations at lower levels. The shared features at lower
levels increase the non-linearity at finer convolutional layers to learn complex mappings.
\subsection{D. Iterative up-and-down sampling:}
However, the previous studies based on back-projection are mostly not deep learning
based and involve some unlearnable operations. To make better use of this
mechanism, Hariset al.exploit iterative up-and-down sampling layers and propose
deep back-projection network (DBPN)\cite{haris2018deep} to mutually connect upsampling layers and
downsampling layers alternately and reconstruct the final HR result using
concatenation of all the intermediately reconstructed HR feature maps. Coupled with
other techniques (e.g., dense connections ), the DBPN wins the championship on the
classical track of NTIRE 2018.
\begin{figure}
\label{sec:fig0}
\begin{minipage}[b]{1\linewidth}
  \centering
  \centerline{\includegraphics[width=8.5cm]{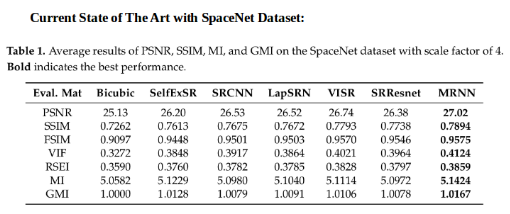}}
  \centerline{Fig 0: Satellite Image Super-Resolution via }\medskip\centerline{'Multi-Scale Residual Deep Neural Network' paper result \cite{xu2019multi}}\medskip
  \end{minipage}
  \end{figure}
 \ref{sec:fig0}
 \begin{figure}
\label{sec:fig4}
\begin{minipage}[b]{1\linewidth}
  \centering
  \centerline{\includegraphics[width=8.5cm]{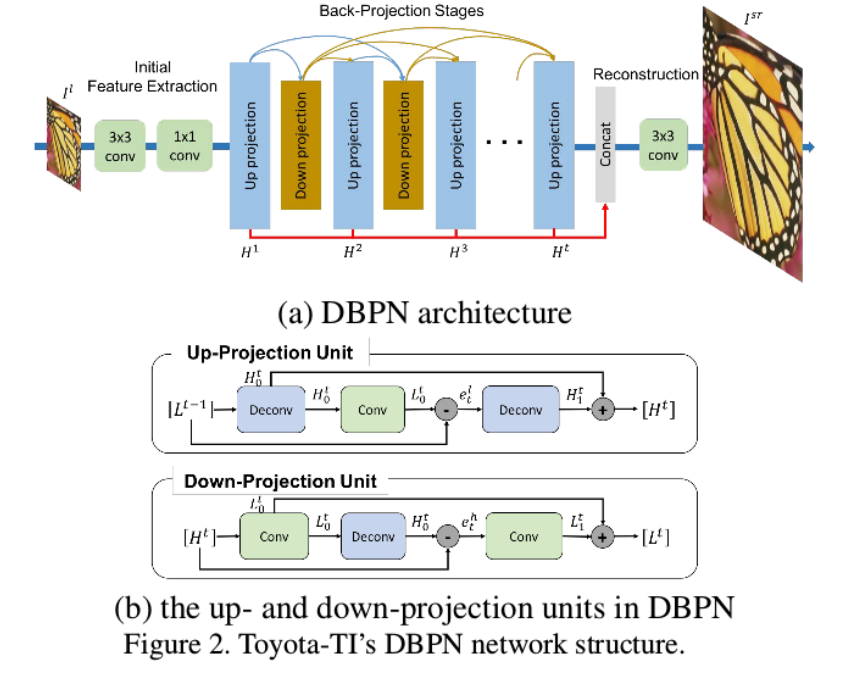}}
  \centerline{Fig 1: DBPN Architectures \cite{haris2018deep}}\medskip
  \end{minipage}
  \end{figure}
 \ref{sec:fig4}
\section{Dataset Description}We used The SpaceNet satellite image dataset for our experiment
The satellite image includes five areas in Rio de Janeiro, Paris, Las Vegas, Shanghai, and Khartoum, which are
collected from DigitalGlobe’s WorldView-2 satellite and published publicly at Amazon. The
complete satellite image of Rio de Janeiro (the spatial resolution is 0.5 m) has the highest resolution
image with 2.8 M ×
2.6 M pixels, and is divided into 6540 non-overlapping HR image patches with 436 × 404 pixels,
and the main contents of interest in the image are buildings and roads.
We randomly sampled 5750 patches from the images out of which 4600 were used for training and
1150 were used for validation.
Low resolution images were prepared by downgrading the original image by a factor of 4 with bicubic interpolation. So, the Low resolution images are of resolution 109 X 101.
\section{Cascaded DBPN:}
\begin{figure}[htb]
\label{sec:fig5}
\begin{minipage}[b]{1\linewidth}
  \centering
  \centerline{\includegraphics[width=8cm]{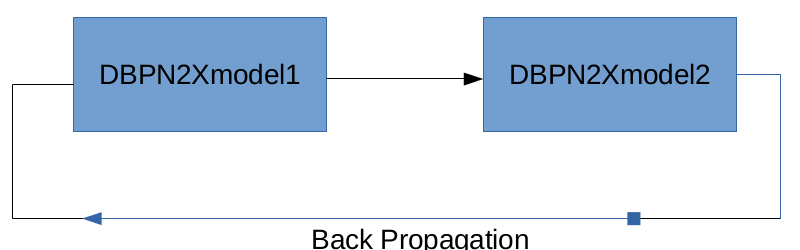}}
  \centerline{Fig 2 :Cascaded DBPN}\medskip
\end{minipage}
\end{figure}
\ref{sec:fig5}
It is just a slight
modification of
already
existing
DBPN models 
that
introduce the
step by step
upsampling. So for getting the image at 4X scale we need to train 2 different DBPN2X models which involve extra computational cost.
\section{Switch Guided Hybrid-Network}
Every deep network has its own pros and cons. The DBPN
model is more suitable for upscaling at two times scale as
for larger scales, it won’t be able to capture the minute detail.
Incorporation of the cascaded structure might be useful here but
this will again add extra computational overhead.
\par Also, some images do not contain significant information, and for this kind of image (only jungles or homogeneous fields), we need not capture any minute object present in the images. We can go with one step upsampling approach or any other shallow methods for this type of image.
\par Hence, Our main objective here is to divide the image set into
different kinds with the help of a classifier, and based on the
switch output, the image patch will be forwarded to a deep or
a shallow network. This will eventually optimize the overall
time overhead. So, we have seen two different aspects of this
problem.
\subsection{ Maintaining computational cost and performance trade-off}
For this, we tried to divide the image patches into two different sets. One set corresponds to the images, giving a better realistic view after performing super-resolution with bicubic interpolation. On the other hand, some images produce distorted output after up-gradation with Bi-cubic interpolation. Below the picture are the examples of these two types of images.
 \par
 \begin{figure}[htb]
\label{sec:fig11}
\begin{minipage}[b]{1\linewidth}
  \centering
  \centerline{\includegraphics[width=8cm]{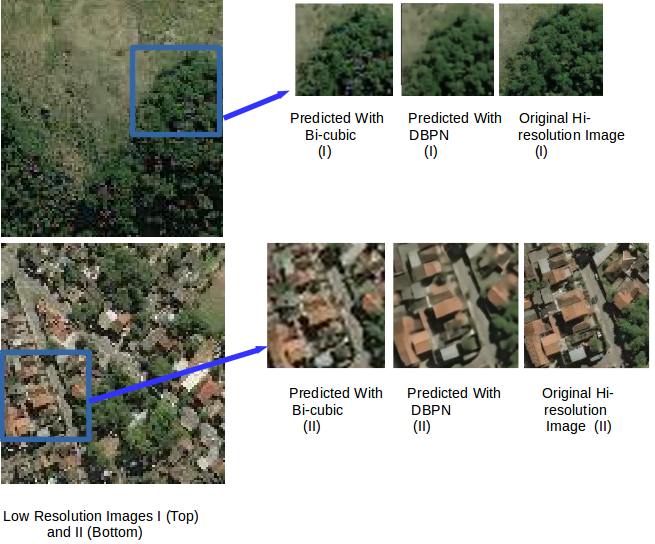}}
  \centerline{Fig 3 Bottom Image contains more information than the top image}\medskip
\end{minipage}
\end{figure}
\ref{sec:fig11}

For the type of image shown in Fig 3 - II(Bottom), Deep Network is more suitable. However, the use of an intense network leads to a high computation time. The computation time of the DBPN network is 1/100th of the time needed for a simple FSRCNN model. FSRCNN is a very shallow network. Hence, it is more effective for real-time super-resolution applications. 
\subsection{The Deep Network or the shallow network focused towards a particular kind of image will enhance the overall performance}

\section{Classification of the individual patches} Here, we argue that classifying the patches based on Output image quality in terms of SSIM is a better choice than classifying the images only based on the edge contents in the input low-resolution image. We have validated our claims with the help of a switching convolution network in the later section.
\par
The Structural SIMilarity (SSIM) index is a method for measuring the similarity between two images. The SSIM index can be viewed as a quality measure of one of the images being compared, provided the other image is regarded as of perfect quality. It is an improved version of the universal image quality index proposed before. Detailed description is given in the following paper:

\par Z. Wang, A. C. Bovik, H. R. Sheikh and E. P. Simoncelli, "Image quality assessment: From error visibility to structural similarity," IEEE Transactions on Image Processing, vol. 13, no. 4, pp. 600-612, Apr. 2004.\cite{wang2004image}
\par To distinguish images for two different types if we had taken the decision based on Edge contents then there are images which have very less edge contents but has some sophisticated background would have been mistakenly classified as easy examples. 
\par
Let's have a look at two such examples:
\begin{figure}[htb]
\label{sec:fig15}
\begin{minipage}[b]{1\linewidth}
  \centering
  \centerline{\includegraphics[width=8cm]{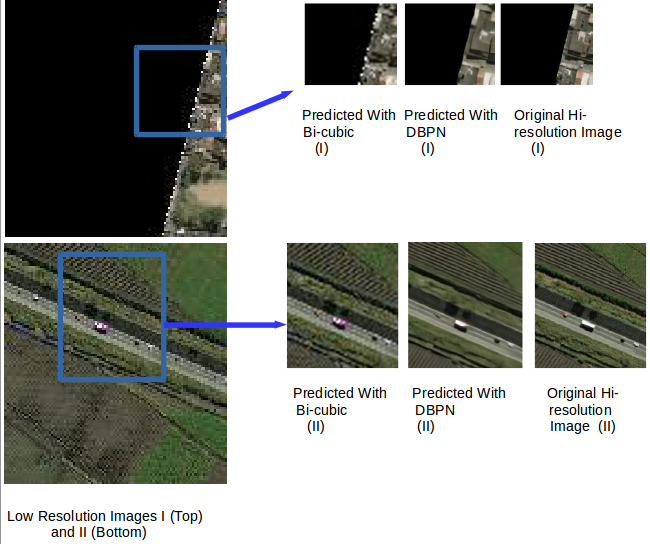}}
  \centerline{Fig 4}\medskip
\end{minipage}
\end{figure}
\ref{sec:fig15}

 We get distorted image after using bi-cubic for the images which attains a poor SSIM value compared to the image which still have better realistic look after performing bi-cubic on the low resolution image. We tried to differentiate the images based on the difference of SSIM achieved from two different method (one is a deep network and another is bicubic) . We have tried with different threshold for the obtained SSIM difference such as .02,,05 and .03.
 
 With .02, almost 50 percent of images in our dataset have been assigned to the 'Difficult set' (the set of patches which contain tiny objects (cars in roads or tank in the rooftop of a building) which can be buildings in urban areas or even the crop field information in the rural areas), and the rest 50 percent to the easier set (The set which doesn't contain any valuable information or objects that has to be present exclusively in the high resolved image). 
\par Next, we tried to train the deep net with only the difficult examples to focus on the images with houses, rooftops, roof tanks, cars, etc. We assume that if we train the network in a way that only gets that difficult example in training, it will give better results than the network trained on the mixed data.
\par Here is another such example in fig 3 top, which has higher entropy than the image in fig 4 top but will not become that much distorted if we use bi-cubic for up-sampling. In Fig 4, top image(I) consists of black regions mostly, but only a small region is densely populated with buildings, etc. So, overall edge contents are comparatively less than the top image in Fig 3(I).
So, the classification, according to the Entropy, could have misclassified these two images.

\section{Methodology:}
\subsection{Dividing the entire dataset into two categories - Easy and Difficult based on the SSIM difference between the predicted images with deep approach and bi-cubic approach }We have tried with different thresholds of SSIM and finally with .02 suited the most if the final super-resolution performance is considered into account.
\begin{figure}[htb]
\label{sec:fig23}
\begin{minipage}[b]{1\linewidth}
  \centering
  \centerline{\includegraphics[width=8cm]{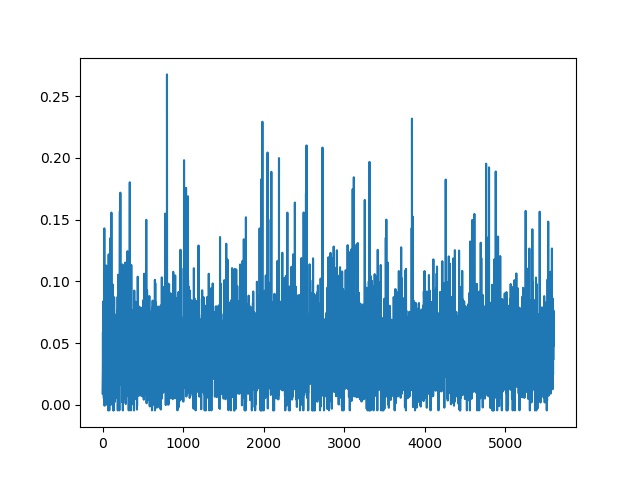}}
  \centerline{Fig 5 :SSIM difference obtained}\medskip
\end{minipage}
\end{figure}
\ref{sec:fig23}
\begin{figure}[htb]
\label{sec:fig23A}
\begin{minipage}[b]{1\linewidth}
  \centering
  \centerline{\includegraphics[width=8cm]{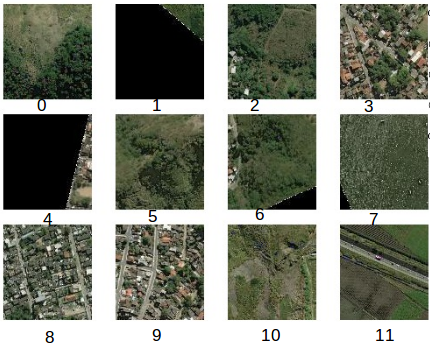}}
  \centerline{Fig 6 :Random Images from Validation Set}\medskip
\end{minipage}
\end{figure}
\ref{sec:fig23A}
\begin{figure}[htb]
\label{sec:fig23B}
\begin{minipage}[b]{1\linewidth}
  \centering
  \centerline{\includegraphics[width=8cm]{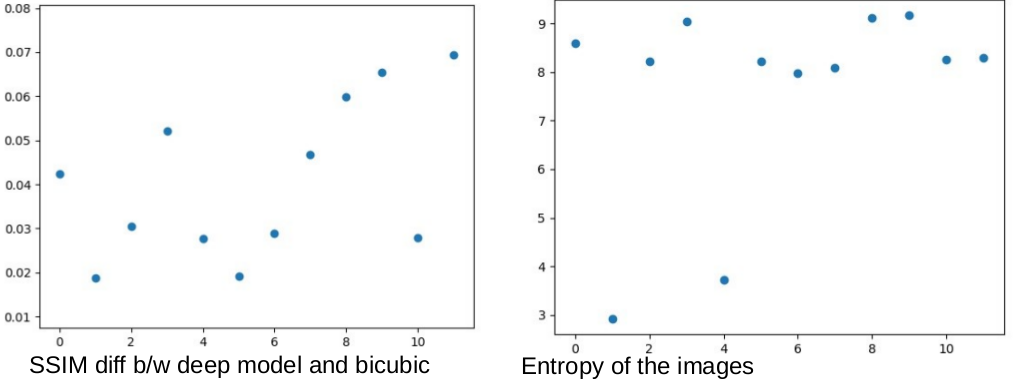}}
  \centerline{Fig 7 :SSIM difference obtained and Entropy}\medskip
  \centerline{for images in fig 6 number from 0 to 11}
\end{minipage}
\end{figure}
\ref{sec:fig23B}
\par Fig in 6 and 7 explains how the SSIM difference(from the image obtained after up-gradation with DBPN and Bi-Cubic interpolation) varies significantly for images containing only green and images which contain houses or other information other than only green and trees or field (could be not entirely but in a small portion as in 4th image). So, SSIM difference captures the two types in a meaningful manner. Entropy alone can not capture this semantic information.
\label{sec:fig23C}
\begin{figure}
\begin{minipage}[b]{1\linewidth}
  \centering
  \centerline{\includegraphics[width=8cm]{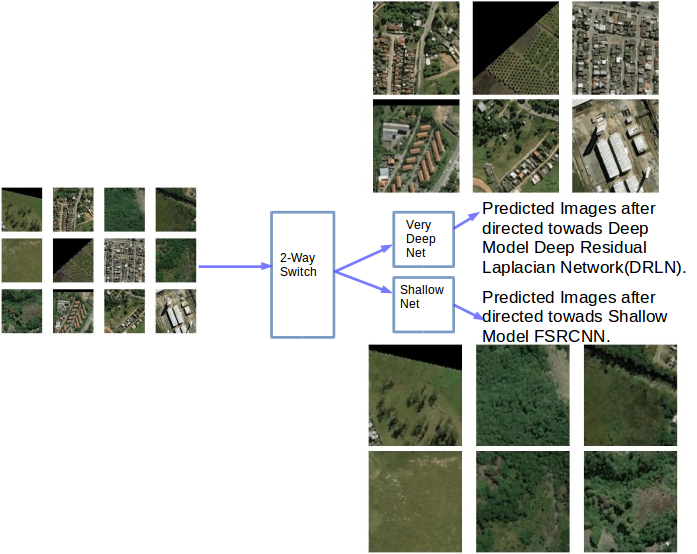}}
  \centerline{Fig 8 Switch Guided Hybrid Network}\medskip
\end{minipage}
\end{figure}
\ref{sec:fig23C}
\subsection{Training of pre-trained VGG16 from the initial convolution layers }

We have trained the VGG 16 network for 3 iterations and obtained 92 percent accuracy in determining the images, whether easy or difficult. 
\subsection{Training of individual Deep Nets based on the outputs from the classifier:}
We have trained FSRCNN with the easy examples tagged by the classifier and DRLN and DBPN with both types of examples, and the overall best PSNR obtained was 31.33 on Easy examples with DBPN and 30.4 with FSRCNN.On the other hand, the overall best PSNR obtained on the difficult set was 24.8 with DBPN and 25.12 with DRLN. 
\section{Verification of our menthods with Coupled training of 5 deep networks and a switch classifier}
Here, the work  \cite{sam2017switching}'Switching Convolutional Neural Network for Crowd Counting' by Deepak Babu, Sam Shiv, Surya R. Venkatesh Babu has motivated us for going with our methodology as we have performed a similar coupled training of 5-way switch and 5 deep networks( DBPN, Cascaded DBPN, FSRCNN, DRLN, and LapSRN). After 2 iterations, all the examples converged to only two deep networks DBPN and DRLN.
 
 \par 
Initially, we trained a pre-trained VGG16 and obtained 72 percent accuracy after 5 iterations with learning rate=.005 based on the labels assigned by the initial models with minimum loss and the numbers assigned to 5 deep networks DBPN, Cascaded DBPN, FSRCNN, DRLN, and LapSRN were 20,3213,1,2282,84 respectively.

\par 
Next, the classifier assigns the 2559 images only to Model 2 and 3041 images to Model 4.
and Overall PSNR achieved as per classifier’s assignment: 28.423

\par Next, we further train the corresponding models with the classifier- assigned labels for one iteration. We obtained the overall best PSNR was 28.56 (taking the best PSNR on an image out of only 2 models).
\par Finally, we train our classifier with the 2 types based on the best PSNR obtained from one of the models and achieve 92 percent accuracy after a single iteration.
\par The classifier assigns 3174 examples to DRLN and 2337 examples to Cascaded DBPN.
\par So, The entire training set containing 5600 examples gets divided into type 1 containing 3174 examples that were assigned to DRLN and type 2 containing 2337 examples that were assigned to cascaded DBPN. 
\par This type 1 set has the most overlapping (80 percent) with our difficult set, and type 2 similarly has a significant overlap with our easy set. Type 1 in the validation set had an overall PSNR of 27.54 and type-2 in the validation set had an overall PSNR of 31, and an overall PSNR on the validation set was 28.49. The validation set consisted of the rest 1400 examples, excluding the training set.
 
\section{Observation}
\begin{figure}[htb]
\label{sec:fig24}
\begin{minipage}[b]{1\linewidth}
  \centering
  \centerline{\includegraphics[width=8cm]{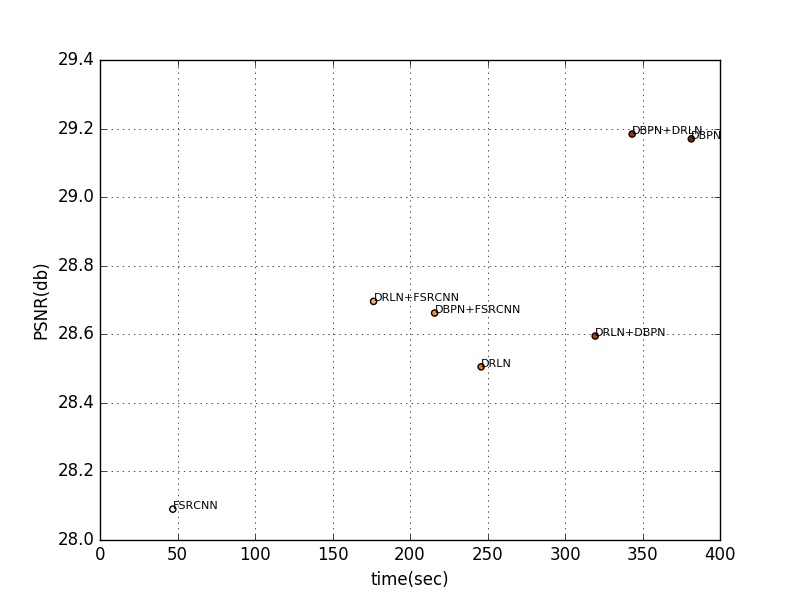}}
  \centerline{Fig:9 time vs PSNR plot for 100 patches}\medskip
\end{minipage}
\end{figure}
\ref{sec:fig24}
\begin{figure}[htb]
\label{sec:fig25}

\begin{minipage}[b]{1\linewidth}
  \centering
  \centerline{\includegraphics[width=8cm]{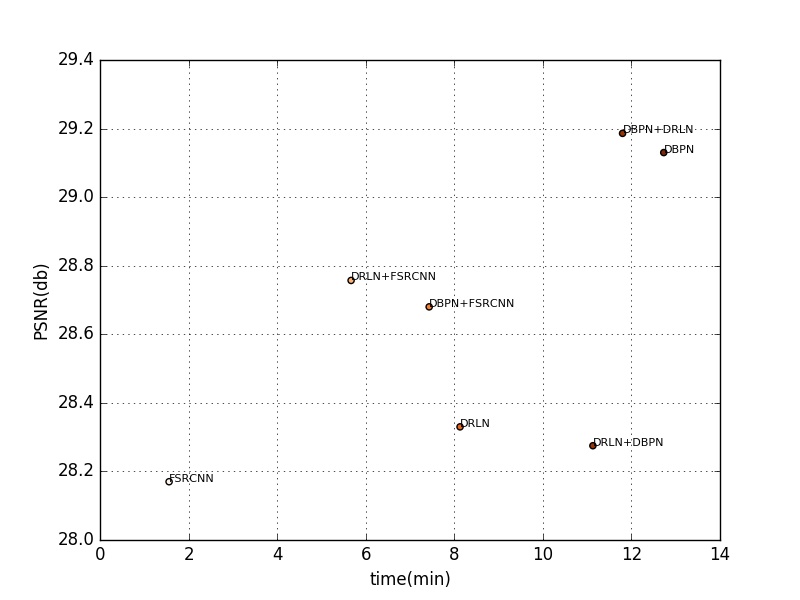}}
  \centerline{Fig:10 time vs PSNR plot for 200 patches}\medskip
\end{minipage}
\end{figure}
\ref{sec:fig25}
\begin{figure}[htb]
\label{sec:fig26}
\begin{minipage}[b]{1\linewidth}
  \centering
  \centerline{\includegraphics[width=8cm]{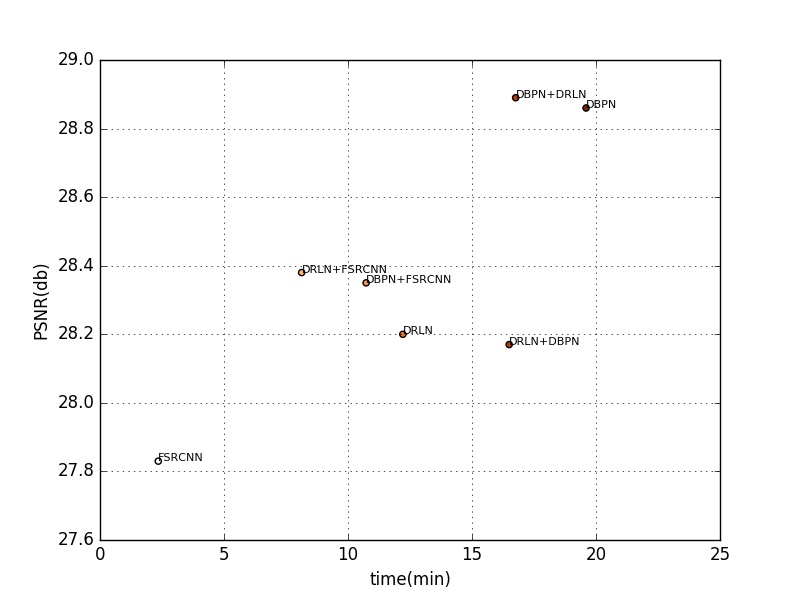}}
  \centerline{Fig:11 time vs PSNR plot for 300 patches}\medskip
\end{minipage}
\end{figure}
\ref{sec:fig26}
\begin{figure}[htb]
\label{sec:fig27}
\begin{minipage}[b]{1\linewidth}
  \centering
  \centerline{\includegraphics[width=8cm]{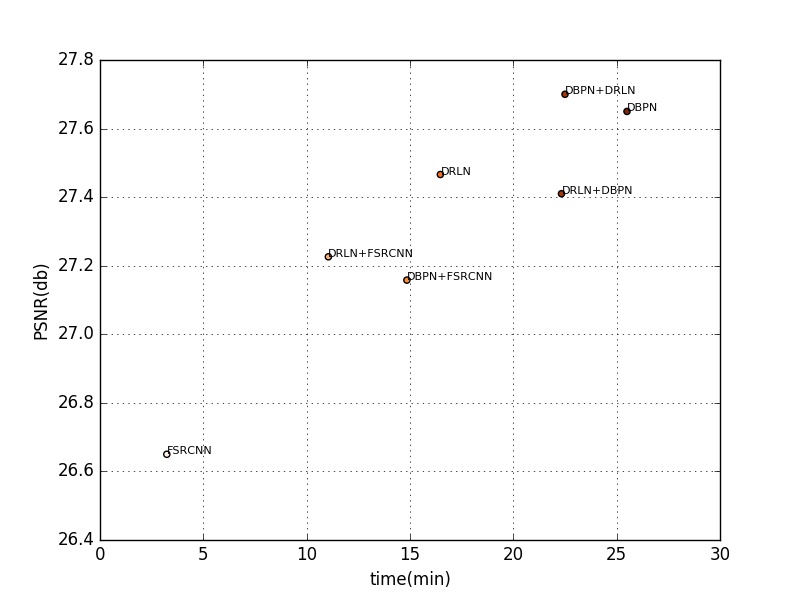}}
  \centerline{Fig:12 time vs PSNR plot for 400 patches}\medskip
\ref{sec:fig27}
\end{minipage}
\end{figure}
\begin{figure}[htb]
\label{sec:fig28}
\begin{minipage}[b]{1\linewidth}
  \centering
  \centerline{\includegraphics[width=8cm]{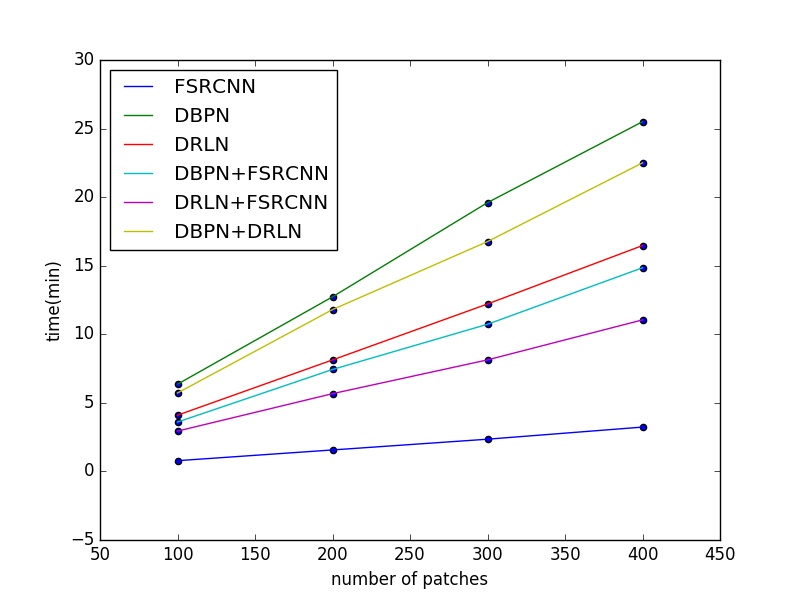}}
  \centerline{Fig:13 number of patches vs time}\medskip
\ref{sec:fig28}
\end{minipage}
\end{figure}
\par
The best performance was obtained with the combination of DBPN (in the absence of cascaded DBPN) on the difficult set and DRLN on the Easy set. Although, using both deep models was computationally expensive. If we had used FSRCNN on the easy set and DRLN on the difficult set, it gave overall better results. The computational cost was not as high as the other two deep models used on both types of images without a classifier (switch) overhead. 
\par We have selected random samples of 100, 200, 300, and 400. We experimented with a different combination of networks and noted down the observation for time and PSNR. 
\section{Result on the entire dataset and Discussion}
The table in Fig 14 contains the PSNR and SSIM obtained on the entire dataset with different models. The combinations of models are obtained with the help of the switch. Although the switching overhead is an extra addition, this approach still takes lesser CPU time (explained with plots in Fig 13). This switch does not classify based on the signal level information but classifies the patches based on high-level semantics. 
\begin{figure}[htb]
\label{sec:fig29}
\begin{minipage}[b]{1\linewidth}
  \centering
  \centerline{\includegraphics[width=8cm]{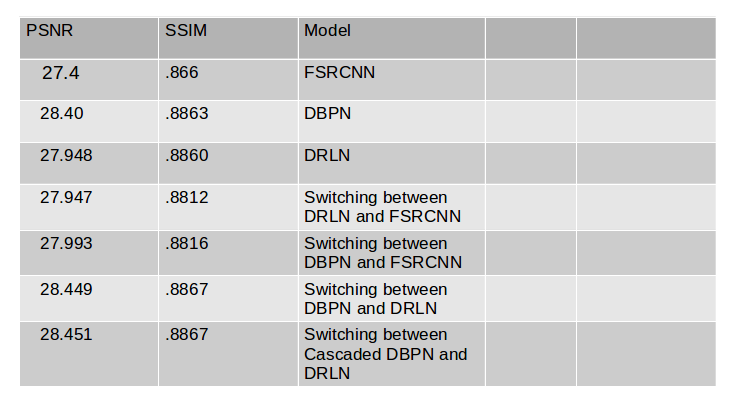}}
  \centerline{Fig:14 SSIM and PSNR obtained on}
  \centerline{the entire dataset}\medskip
\ref{sec:fig29}
\end{minipage}
\end{figure}
\par
\section{Conclusions}
As we can see (in Fig 13) that with the increase of the number of patches, the time increase rate is minimized if we had used the combination of DRLN and FSRCNN or DBPN and FSRCNN instead of using DRLN or DBPN altogether. So, the switch is smart enough to determine the respective network (either shallow or deep) in a manner that can lower the computational time for a large number of patches.

\par
Also, with the switch determined Hybrid Network (a combination of DRLN on Difficult and DBPN on Easy set) gives a rise in the overall performance from 28.40 (if we had used only DBPN) to 28.45 if we had used DBPN on easy set and DRLN on difficult set.
\section{Future Work}
We can implement this approach on a different task. The application can be a small font extraction from the low-resolution image. We can use a similar approach of classifying the examples into two categories and send the images either to a deep network or to a shallow network depending on the font resolution and edge contents in the image.


\bibliography{root.blb}
\bibliographystyle{plain}
\end{document}